\documentclass[sigconf,10pt, screen, nonacm]{acmart}

\usepackage{amsmath,amsfonts}
\usepackage[noend]{algorithmic}
\usepackage{lipsum}
\usepackage{graphicx}
\usepackage{listings}
\usepackage{algorithm}
\usepackage{enumitem}
\usepackage{setspace}
\usepackage{graphbox}
\usepackage{multirow}
\usepackage{subcaption} %
\usepackage[pdf]{graphviz}
\usepackage[font=small,skip=3pt]{caption}
\usepackage{textcomp}
\usepackage{url}

\usepackage{acmart-taps}

\NewDocumentCommand{\code}{v}{%
\texttt{\textcolor{black}{#1}}%
}

\definecolor{codegreen}{rgb}{0,0.6,0}
\definecolor{codegray}{rgb}{0.5,0.5,0.5}
\definecolor{codepurple}{rgb}{0.58,0,0.82}
\definecolor{backcolour}{rgb}{0.95,0.95,0.92}

  {\list{}{\leftmargin=0.0in\rightmargin=0.0in}  \item[]  }%
  {\endlist}

\lstdefinestyle{mystyle}{
    backgroundcolor=\color{backcolour},
    commentstyle=\color{codegreen},
    keywordstyle=\color{magenta},
    numberstyle=\tiny\color{codegray},
    stringstyle=\color{codepurple},
    basicstyle=\ttfamily\footnotesize,
    breakatwhitespace=false,
    breaklines=true,                 
    captionpos=b,                    
    keepspaces=true,                 
    numbers=left,                    
    numbersep=5pt,                  
    showspaces=false,                
    showstringspaces=false,
    showtabs=false,                  
    tabsize=2
}
\copyrightyear{2025}
\acmYear{2025}
\setcopyright{cc}
\setcctype{by}
\acmConference[ICS '25]{2025 International Conference on Supercomputing}{June 8--11, 2025}{Salt Lake City, UT, USA}
\acmBooktitle{2025 International Conference on Supercomputing (ICS '25), June 8--11, 2025, Salt Lake City, UT, USA}
\acmDOI{10.1145/3721145.3734530}
\acmISBN{979-8-4007-1537-2/2025/06}

\acmConference[ICS'2025]{The 39th ACM International Conference on
  Supercomputing}{June 9--11, 2025}{Salt Lake City, USA}

\begin{document}

\title{EDAN: Towards Understanding Memory Parallelism and Latency Sensitivity in HPC}

\author{Siyuan Shen}
\affiliation{%
  \institution{ETH Z\"urich}
  \country{Switzerland}
}
\email{siyuan.shen@inf.ethz.ch}

\author{Mikhail Khalilov}
\affiliation{%
  \institution{ETH Z\"urich}
  \country{Switzerland}
  }
\email{mikhail.khalilov@inf.ethz.ch}

\author{Lukas Gianinazzi}
\affiliation{%
  \institution{ETH Z\"urich}
  \country{Switzerland}
}
\email{lukas.gianinazzi@inf.ethz.ch}

\author{Timo Schneider}
\affiliation{%
 \institution{ETH Z\"urich}
 \country{Switzerland}
}
\email{timo.schneider@inf.ethz.ch}

\author{Marcin Chrapek}
\affiliation{%
 \institution{ETH Z\"urich}
  \country{Switzerland}
}
\email{marcin.chrapek@inf.ethz.ch}

\author{Jai Dayal}
\affiliation{%
  \institution{Cerebras Systems}
  \country{USA}
}
\email{jai.dayal@cerebras.net}

\author{Manisha Gajbe}
\affiliation{%
  \institution{Not Affiliated}
  \country{USA}
}
\email{manisha.gajbe@gmail.com}

\author{Robert Wisniewski}
\affiliation{%
  \institution{Hewlett Packard Enterprise}
  \country{USA}
}
\email{bobww123@gmail.com}

\author{Torsten Hoefler}
\affiliation{%
  \institution{ETH Z\"urich}
  \country{Switzerland}
}
\email{torsten.hoefler@inf.ethz.ch}

\renewcommand{\shortauthors}{Shen et al.}

\begin{abstract}
Resource disaggregation is a promising technique for improving the efficiency of large-scale computing systems. However, this comes at the cost of increased memory access latency due to the need to rely on the network fabric to transfer data between remote nodes. As such, it is crucial to ascertain an application's memory latency sensitivity to minimize the overall performance impact. Existing tools for measuring memory latency sensitivity often rely on custom ad-hoc hardware or cycle-accurate simulators, which can be inflexible and time-consuming. To address this, we present \emph{EDAN} (\underline{E}xecution \underline{D}AG \underline{An}alyzer), a novel performance analysis tool that leverages an application's runtime instruction trace to generate its corresponding execution DAG. This approach allows us to estimate the latency sensitivity of sequential programs and investigate the impact of different hardware configurations. EDAN not only provides us with the capability of calculating the theoretical bounds for performance metrics, but it also helps us gain insight into the memory-level parallelism inherent to HPC applications. We apply EDAN to applications and benchmarks such as PolyBench, HPCG, and LULESH to unveil the characteristics of their intrinsic memory-level parallelism and latency sensitivity.
\end{abstract}



\pagenumbering{gobble}
\maketitle

\section{Introduction}

Modern communication networks exhibit exponentially increasing bandwidths, exemplified by the doubling of Ethernet switch rates every two years~\cite{datacenter_ethernet_hoefler}. These higher bandwidths are achieved through higher frequencies and complex signaling (e.g., PAM4), resulting in higher bit error rates in transceivers. To address this, stronger and more complex forward error correction (FEC) mechanisms, such as those in the upcoming 800G and 1.6T IEEE P802.3df specification~\cite{ieee_dambrosia}, are used, significantly increasing processing latency. Fast FEC implementations today can have latencies as low as 50ns, but future FECs may increase this latency by 100ns or more, leading to per-link latency of several hundred nanoseconds~\cite{datacenter_ethernet_hoefler}. This trend continues across modern networks, with higher bandwidths at the expense of higher latencies.

Resource disaggregation is a promising technique that has been recently explored in both the fields of High-Performance Computing (HPC) and datacenter designs. It challenges the traditional monolithic server architecture by separating heterogeneous resources into discrete units connected by a high-speed network. Not only does this approach allow flexible and dynamic provisioning of resources to better match various application requirements, but it also minimizes the idle time of expensive resources, such as accelerators and CPUs, which greatly reduces the cost and energy consumption in a large system \cite{intra_rack_resource_disaggregation_michelogiannakis}. While memory disaggregation networks opt for lower-latency protocols and weaker FEC protection, they follow the same general trend as Ethernet: higher bandwidth will likely cause higher latency. Thus, the ratio of bandwidth to latency will worsen in the coming generations.

Memory disaggregation is prevalent amongst all the resource aggregation systems \cite{memory_disaggregation_liu, memory_underutilization_peng, disaggregated_memory_lim, software_resource_disaggregation_copik}, as it increases the memory utilization across datacenters and helps boost the scalability of memory-intensive applications, such as data-processing frameworks and HPC applications \cite{evaluating_hardware_memory_disaggregation_patke}. However, growing latencies may limit or even reduce their efficiency because data access relies on the network fabric~\cite{evaluating_hardware_memory_disaggregation_patke}. Gao et al. \cite{network_requirements_gao} show that additional latency reduces the performance of data-intensive applications regardless of the network bandwidth. In essence, the more sensitive an application is toward memory latency, the more noticeable its performance degradation will be. To this end, it is crucial to ascertain the memory latency sensitivity and tolerance of applications so that resource allocations and system design can be done in a way that minimizes the overall performance impact.

Measuring memory latency sensitivity, however, is a complex topic. It generally involves artificially injecting latency into memory accesses and recording application runtime under varying degrees of additional latency. As exemplified by the works of Patke et al. \cite{evaluating_hardware_memory_disaggregation_patke} and Domke et al. \cite{locus_of_performance_domke}, one has the option to depend either on some custom ad-hoc hardware that is inflexible and difficult to acquire or cycle-accurate simulators that are extremely time-consuming.

\begin{figure*}[!t]
    \centering
    \includegraphics[width=\textwidth]{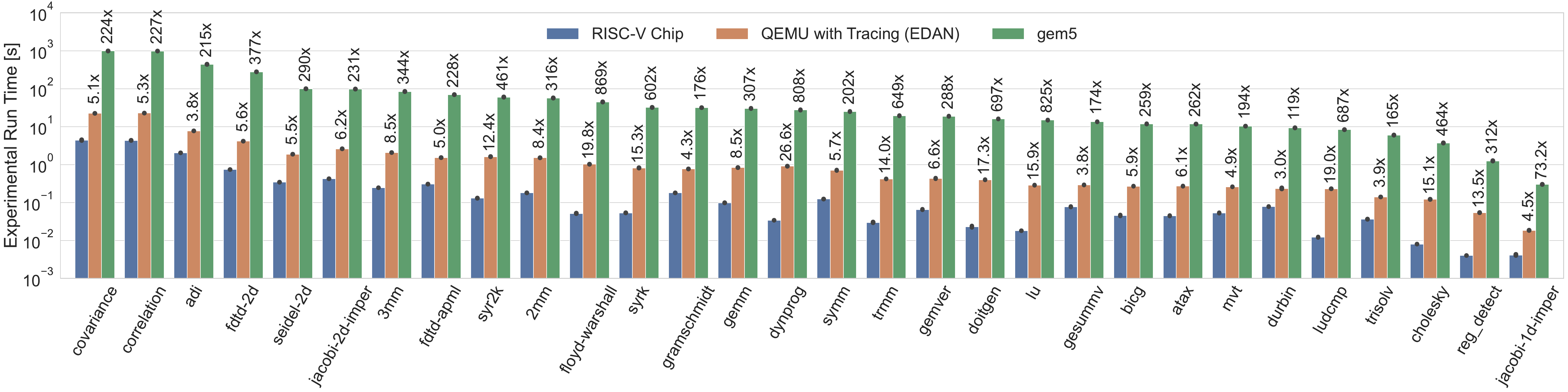}
    \caption{Simulation time of Polybench kernels (small size) using QEMU emulation with instruction tracing (EDAN), and gem5 cycle-approximate simulation. Runtime on a RISC-V chip is used as the baseline for slowdown calculations.}
    \label{fig:motivation}
\end{figure*}

To address this issue, we introduce \emph{EDAN} (\underline{E}xecution \underline{D}AG \underline{An}alyzer), a novel performance analysis tool that leverages a program's runtime instruction trace to generate its corresponding execution DAG (eDAG). This approach exposes true instruction dependencies and allows us to estimate an application's latency sensitivity in fine detail using a CPU and cache model. Unlike other methods, EDAN requires only one execution of the program to automatically generate the eDAG, enabling efficient investigation of different hardware configurations (e.g., cache sizes, memory issue slots). EDAN empowers programmers to prioritize memory latency tolerance in algorithm design and provides valuable insights for hardware architects on the impact of architectural parameters. In the context of HPC, EDAN becomes essential for memory-intensive tasks, identifying code sections with high memory intensity and optimizing parallel execution and data locality for enhanced performance.

Under the assumption of an idealized computational model, EDAN not only provides us with the capability of calculating the theoretical bounds for performance metrics such as bandwidth utilization and memory latency sensitivity, but it also helps us gain preliminary insights into the memory-level parallelism inherent to a range of applications. This, in turn, can guide design decisions about architectural parameters, such as the number of issue slots and cache sizes. We apply EDAN to applications and benchmarks such as PolyBench, HPCG, and LULESH to shed some light on their memory-level parallelism and latency sensitivity.

The primary contributions of our paper are as follows:

\begin{itemize}
\item We develop EDAN, an experimental tool for theoretical performance analysis based on execution DAGs.
\item We define a new memory cost model inspired by Brent's lemma, which defines upper and lower bounds for the memory access cost of an eDAG based on the number of memory issue slots. From this model, we then derive two performance metrics that quantify the memory latency sensitivity of an application.
\item We demonstrate the effectiveness of EDAN by applying it to several HPC applications and benchmarks, and present the insights from this investigation.
\end{itemize}

\subsection{Motivation}

To assess the memory latency sensitivity of an application, state-of-the-art cycle-approximate simulators such as gem5 are commonly used. Despite its flexibility and relative accuracy, one significant drawback is its simulation speed. As addressed in \cite{gem5_binkert}, compared with translation-based simulators such as QEMU \cite{qemu_bellard}, gem5 is significantly slower. This was demonstrated in \cite{locus_of_performance_domke}, which claimed to be ``the largest cycle-accurate simulations" ever conducted with research-driven gem5. As the authors stated, the benchmarks alone took multiple months to run, and even then some were still missing due to gem5-related issues or exceeding the time limit of the simulation. Despite the complexity and scale of their experiments, it is evident that gem5 lacks scalability.

We tested the slowdown of gem5 (version 22.1) using PolyBench-C (version 3.20), cross-compiled into RISC-V binaries. We ran the benchmarks in three environments: \textbf{(i)} native RISC-V chip, \textbf{(ii)} QEMU user-mode emulation with a custom instruction tracing plugin, and \textbf{(iii)} gem5. The RISC-V board used was a StarFive VisionFive with 2 CPUs and 8GB of memory. The server for QEMU and gem5 had an AMD Ryzen 5 CPU and 16GB of RAM. The gem5 configuration included SE mode, 1 GHz RiscvO3CPU with 16GB DRAM (50ns latency), 16kB L1i, 64kB L1d caches, and 256kB L2 cache. In Fig \ref{fig:motivation}, gem5 showed slowdowns ranging from $100\times$ to $900\times$, while our plugin was on average only $5\times$ to $10\times$ slower than the baseline. %
This discrepancy highlights the scalability issues of using gem5 for large HPC applications or parameter sweeps. In contrast, EDAN uses QEMU for tracing, which is an order of magnitude faster than gem5.

To generate DAGs for a program, one may propose to exploit memory traces or MPI traces, yet neither is sufficient in this scenario. A pure memory trace does not contain dependency information between memory accesses, and an MPI trace is too coarse-grained for exposing memory-level parallelism. Therefore, our approach is necessary as it allows us to accurately identify data dependencies between instructions in the most fine-grained manner.

\section{Background}

\subsection{Execution DAG (eDAG)}
\label{sec:edag}

\begin{figure}[!htp]
    \centering
    \includegraphics[width=\linewidth]{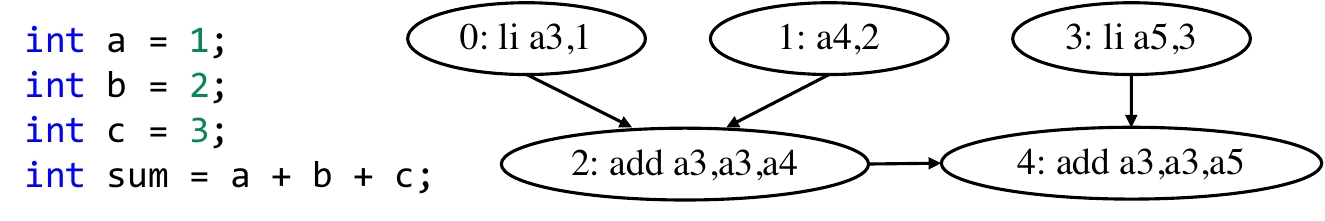}
    \caption{A simple C program calculating the sum of 3 variables and its corresponding eDAG.}
    \label{fig:eDAG-example}
\end{figure}

An execution DAG (eDAG) is a way to represent the data dependencies between computations. While they share similarities with the computational DAG commonly discussed in the literature~ \cite{pebbles_graphs_kwasniewski, red_blue_pebbling_revisited_kwasniewski, io_lower_bounds_zhang}, eDAGs are distinct in that they are generated from execution traces instead of being derived from analyzing programs' computation patterns. Formally, an eDAG can be expressed as a directed graph $G = (V, E)$. $V$ represents the set of instructions in a program and edges $E \subseteq (V \times V)$ denote the set of directed edges defining the data dependencies between instructions \cite{survey_on_parallel_computing_navarro, introduction_to_algorithms_cormen, DAG_scheduling_zhao}. Fig \ref{fig:eDAG-example} provides an example of a trivial C program in which three variables \texttt{a}, \texttt{b}, and \texttt{c} are initialized, and added together to another variable \texttt{sum}.

\subsection{DAG-based Performance Analysis}
\label{sec:dag-analysis}

$T_1$ is the total time needed to execute all instructions in the program with one processor \cite{introduction_to_algorithms_cormen}. Mathematically, $T_1 = \sum_{v \in V} t(v)$, where $t(v)$ gives the execution time of vertex $v$. If all vertices have unit cost, $T_1$ equals the total number of vertices in the eDAG. The depth of an eDAG $T_{\infty}$, also known as span, is the shortest time required to execute all instructions with unlimited processors. It is the aggregate execution time of vertices along the longest path, which is the critical path of the eDAG. The depth is $T_{\infty} = \max_{\pi} T(\pi)$, where $\pi$ denotes a sequence of vertices $v_1, v_2, \ldots, v_k$ with $v_1$ and $v_k$ as input and output vertices, respectively, and $T(\pi) = \sum_{v \in \pi} t(v)$. The degree of parallelism is the ratio between $T_1$ and $T_\infty$, representing the average number of vertices that can be executed concurrently at each step along the critical path. A higher degree of parallelism means more tasks can be executed simultaneously, potentially leading to faster overall program execution time. The \emph{work law} states $T_p \geq \frac{T_1}{p}$, where $p$ is the number of available processors, and $T_p$ is the execution time of the program. The \emph{span law} states that $T_p \geq T_\infty$. Together, they define the lower bound of $T_p$ as $T_p \geq \max \{\frac{T_1}{p}, T_\infty\}$. Brent's lemma, with $T_1$, $T_\infty$, $p$ processors, and a greedy scheduler, states that the execution time of a program $T_p$ is upper bounded by $\frac{T_1 - T_\infty}{p} + T_\infty$ \cite{parallel_evaluation_brent, brent_theorem_gustafson}.

\section{eDAG Analyzer Toolchain}
\label{sec:edan}
\begin{figure}[!htp]
    \centering
    \includegraphics[width=1\linewidth]{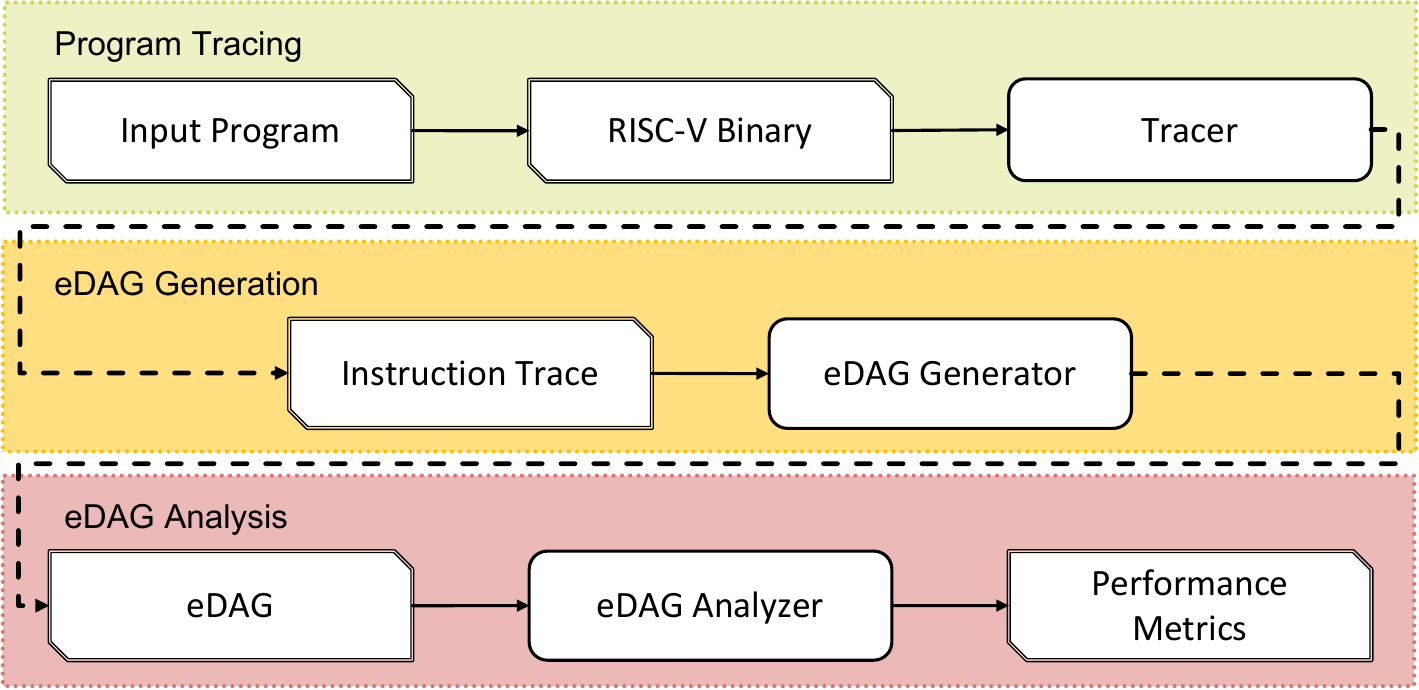}
    \caption{High-level overview of the EDAN toolchain.}
    \label{fig:eDAG-pipeline}
\end{figure}

Fig \ref{fig:eDAG-pipeline} illustrates the overall structure and the elements involved in the workflow of the EDAN toolchain. As one can see from the colored blocks, it is composed of three main stages, whose respective functionalities are tracing programs, generating eDAGs based on the collected trace, and producing relevant performance metrics from eDAGs. The principal components and design choices in each stage are discussed in detail in the following sections.

We chose RISC-V as the target ISA for two reasons. Firstly, it is relatively simple with fewer and less complex instructions, speeding up development and simplifying the parser~\cite{riscv_manual_waterman}. Secondly, RISC-V has garnered significant interest from academia and industry, leading to the development of numerous new extensions and hardware support~\cite{instruction_sets_should_be_free_asanovic, comparative_survey_dorflinger, quark_askarihemmat, indirection_stream_semantic_scheffler}. Our aim is to contribute to the open-source RISC-V community and ecosystem through EDAN's introduction. EDAN's modular design allows for easy incorporation of support for other ISAs without affecting the core functionality of the toolchain.

\subsection{Program Tracing}

The primary goal of the first stage is to obtain a trace of every assembly instruction that has been executed in a program. To start, we take the source code of an arbitrary application and compile it to RISC-V binary. In order to achieve this, we primarily leveraged the RISC-V GNU Toolchain (GCC version 12.2.0) \cite{riscv_gnu_toolchain}, considering that many users may not have access to dedicated hardware supporting the RISC-V ISA. To ensure optimal performance, all programs mentioned in this paper are compiled using \textbf{O3} optimization.

\begin{figure}[!htp]
\centering
\begin{lstlisting}[
    language=C,
    backgroundcolor=\color{white},
    basicstyle=\fontsize{7}{7}\ttfamily,
    xleftmargin=0.5\linewidth,
]
#define N 4
int kernel(int *arr, int n)
{
  int i, sum = 0;
  // Perform summation
  for (i = 0; i < n; ++i)
      sum += arr[i];
  return sum;
}
\end{lstlisting}
\caption{Kernel in C that sums all elements in an array.}
\label{lst:summation-example-source}
\end{figure}
\begin{figure}
\centering
\begin{lstlisting}[backgroundcolor=\color{white}, basicstyle=\fontsize{7}{7}\ttfamily, xleftmargin=0.5\linewidth]
add a3,a0,a1
mv a0,zero
lw a4,0(a5);0x40080290
addi a5,a5,4
addw a0,a0,a4
bne a3,a5,-6
lw a4,0(a5);0x40080294
addi a5,a5,4
addw a0,a0,a4
\end{lstlisting}
\caption{Section of the trace from the summation kernel.}
\label{lst:summation-trace}
\end{figure}

\begin{figure*}
\begin{subfigure}{.49\textwidth}
\centering
\includegraphics[height=2.3cm]{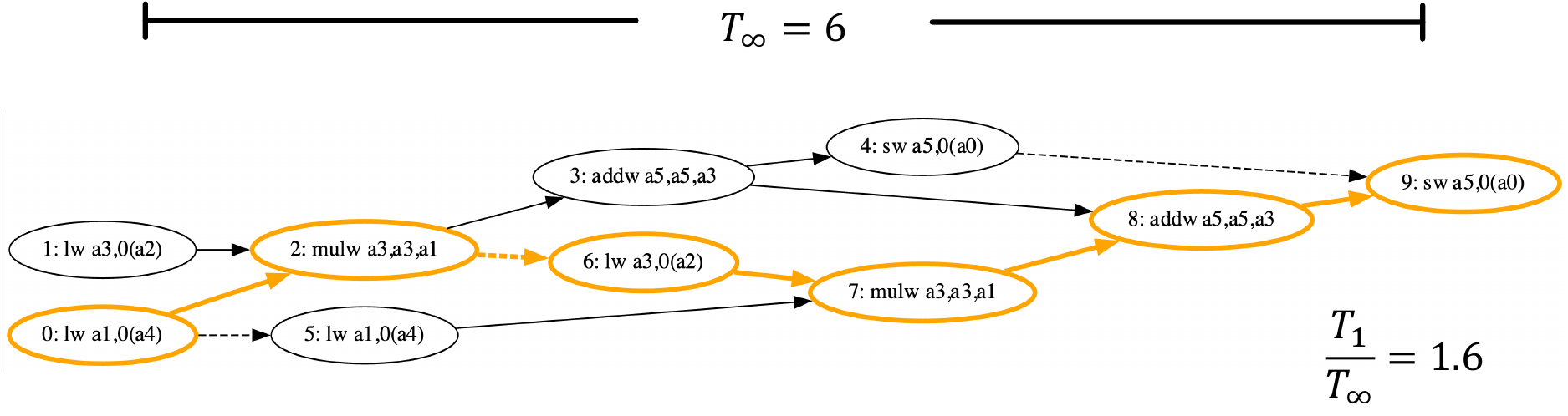}
\caption{Example eDAG where non-true dependencies are dashed arrows.}
\label{fig:false-dep-subfigure}
\end{subfigure}
\hspace{0.01\textwidth}%
\begin{subfigure}{.49\textwidth}
\centering
\includegraphics[height=2.3cm]{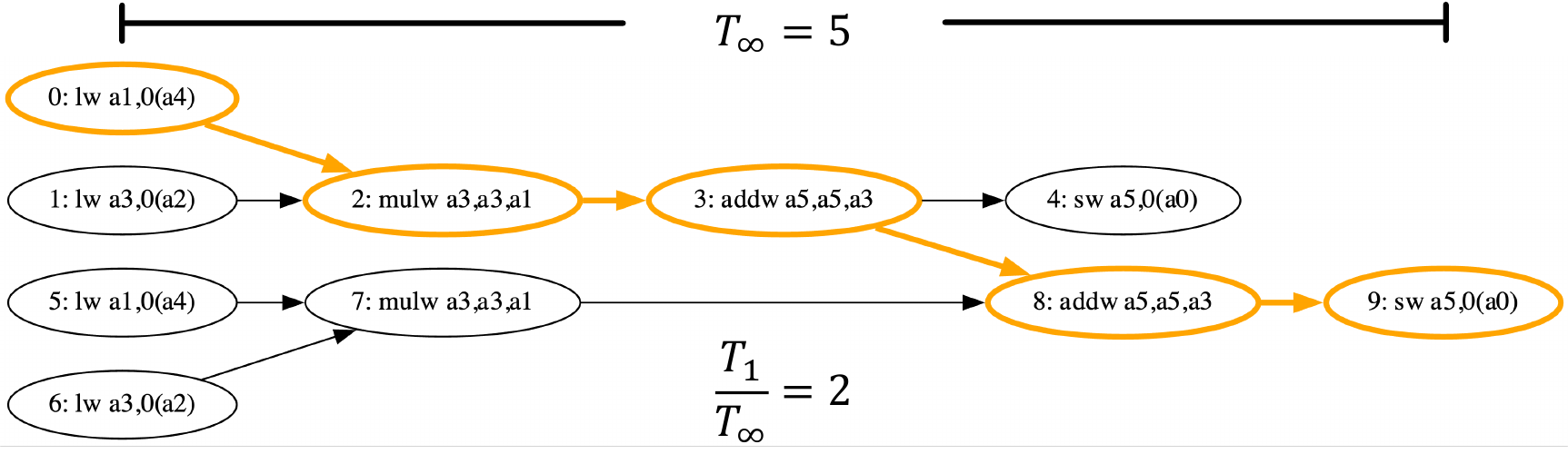}
\caption{The same eDAG where only true dependencies are present.}
\label{fig:false-dep-corrected-subfigure}
\end{subfigure}
\caption{Removing non-true dependencies can help reduce the depth of the eDAG and expose potential parallelism. A critical path in both graphs is highlighted.}
\label{fig:false-dep-example}
\end{figure*}

\subsubsection{Tracer}

Many tools can be employed to trace programs, including \emph{perf} \cite{perf_wiki} or \emph{gdb} \cite{gdb_manual}.  Nonetheless, they are either too slow or do not allow the trace output to be customized easily. To this end, we chose to utilize the Tiny Code Generator (TCG) plugin in QEMU (version 7.2.91) user mode \cite{tcg_plugin_bennee} as the core of EDAN's tracer. This approach has several benefits. Firstly, QEMU under user mode is exceptionally fast as TCG translates target instructions and syscalls to be host-compatible without emulating the OS kernel or hardware. Secondly, TCG plugins are C programs that access the runtime information and interact with QEMU via APIs. Hence, by writing our own TCG plugin and modifying parts of the disassembler, we can easily tailor the output to our desired format and maximize the performance of program tracing. Lastly, unlike ISA-specific emulators such as rv8 \cite{rv8_clark} and banshee \cite{banshee_riedel}, similar TCG plugins can be attached to QEMU emulators with different target ISAs, enabling assembly traces of various ISAs to be collected.
The tracer plugin also provides the flexibility to specify functions for tracing or exclusion. This approach records only instructions from crucial functions, ignoring irrelevant calls to the runtime library, reducing the overhead and noise. In this case, adding compiler flags like \texttt{-g} or \texttt{-fno-inline} is necessary to ensure corresponding symbols are accessible during tracing.

Fig \ref{lst:summation-example-source} shows a code that traverses through an integer array \texttt{arr} of size \texttt{n} and returns the sum of all of its items. It will serve as a running example throughout the paper and be referred to as the summation kernel. Fig \ref{lst:summation-trace}, on the other hand, presents one section of the trace that was obtained through the execution of the summation kernel. The instruction trace contains two columns of data separated by semicolons. The first column shows the assembly instruction, and the last column, which is not always present, denotes the virtual address of the data of a memory access operation.

\subsection{eDAG Generation}

 After program tracing, the next step is to parse the trace and generate an appropriate eDAG. To achieve this, we developed a trace parser and eDAG generator written in Python.

The principal ideas behind the eDAG generator are presented by the Python-style pseudocode in Algorithm \ref{alg:edag-generator}.

The eDAG generator is presented through Python-style pseudocode in Algorithm \ref{alg:edag-generator}. It processes the trace file, extracting instructions and data addresses. The cache model simulates cache hits, and the instruction cost model computes the computation cost of each instruction. These models provide essential information for computing performance metrics from eDAGs (Section \ref{sec:edag}). The core algorithm (lines 10 to 15) establishes dependencies between vertices, ensuring correct identification of data dependencies during eDAG construction. The \texttt{generate\_vertex()} function abstracts ISA-specific functionalities, simplifying the incorporation of more ISAs into EDAN.

One potential restriction of applying the cache model to the memory addresses according to the sequential order in the trace is that when $N$ memory accesses are executed, there are in reality $N!$ ways to order them. Distinctive orderings will result in dissimilar cache miss rates. Therefore, all topological sortings of memory access vertices should be considered in theory. Nonetheless, that would be computationally intractable, and thus we decided to only follow one specific ordering of the memory accesses.

\begin{algorithm}[!htp]
\begin{algorithmic}[1]
\REQUIRE Trace file $trace$, Cache model $cache$ with parameters $\theta$, Instruction cost model $t$ with parameters $\phi$
\STATE Initialize new $eDAG$ object
\STATE Initialize a dictionary $curr\_vs$
\FOR{each $line$ in trace}
    \STATE $insn$, $data\_addr$ = $line.split($`;'$)$
    \STATE $v = generate\_vertex(insn, data\_addr)$
    \IF{$data\_addr$ is not None}
        \STATE $v.cache\_hit = cache.get(data\_addr)$
    \ENDIF
    \STATE $v.time = t.get\_cost(v)$
    \STATE $eDAG.add\_vertex(v)$
    \FOR{each $val$ in $v.dep\_vals$}
        \STATE $dep\_v = curr\_vs[val]$
        \IF{$dep\_v$ exists}
            \STATE $eDAG.add\_edge(dep\_v, v)$
        \ENDIF
    \ENDFOR
    \FOR{each $target$ in $v.targets$}
        \STATE $curr\_vs[target] = v$
    \ENDFOR
\ENDFOR
\RETURN $eDAG$
\end{algorithmic}
\caption{Pseudocode for generating eDAGs}
\label{alg:edag-generator}
\end{algorithm}

\subsubsection{Exposing Potential Parallelism}
\label{sec:exposing-potential-parallelism}
There are four categories of data dependencies, which include true dependencies (read-after-write dependencies or RAW), anti-dependencies (write-after-read dependencies or WAR), output dependencies (write-after-write dependencies or WAW), and input dependencies (read-after-read or RAR)~\cite{discovery_zhen}. We noticed that considering non-true data dependencies in eDAGs greatly hinders the discovery of potential instruction-level parallelism that is intrinsic to an application~\cite{out_of_order_mutlu}. This can be primarily attributed to the fact that in a realistic microarchitecture, only a limited number of registers are available. Hence, false dependencies, especially WAW, can be seen as introduced by the register allocation algorithm in a compiler~\cite{evaluation_of_algorithms_jahnichen} as a means to cope with this constraint.

To demonstrate how removing non-true dependencies exposes potential instruction-level parallelism from the trace of a purely sequential program, we present Fig \ref{fig:false-dep-example}. The two subfigures show a segment of an eDAG generated from a matrix multiplication kernel. The difference is that in Fig \ref{fig:false-dep-subfigure}, both WAW and RAW dependencies are kept,
while in Fig \ref{fig:false-dep-corrected-subfigure} WAW dependencies are removed. Assuming that all vertices have unit cost, the work $T_1$ in both cases is equal to 10, yet in the first scenario, $T_\infty$ is 6 whereas $T_\infty$ is 5 in the second scenario. In this specific example, as vertex 6 no longer needs to wait for vertex 2 to free register \texttt{a3}, all load instructions can be executed at the same time. Thus, by ignoring false data dependencies, we increased the average degree of parallelism from 1.6 to 2.

One limitation of our current approach is that it cannot fully uncover the potential instruction-level parallelism when register spilling happens. To be more precise, depending on the register pressure of the ISA and the design of the compiler, the values of some variables will be spilled to the main memory and stored back throughout the execution of the program \cite{register_allocation_chaitin}. This process creates additional dependencies between instructions and decreases the maximum degree of parallelism the program is capable of achieving given enough registers.

\begin{figure}[!htp]
    \centering
    \includegraphics[height=3.5cm]{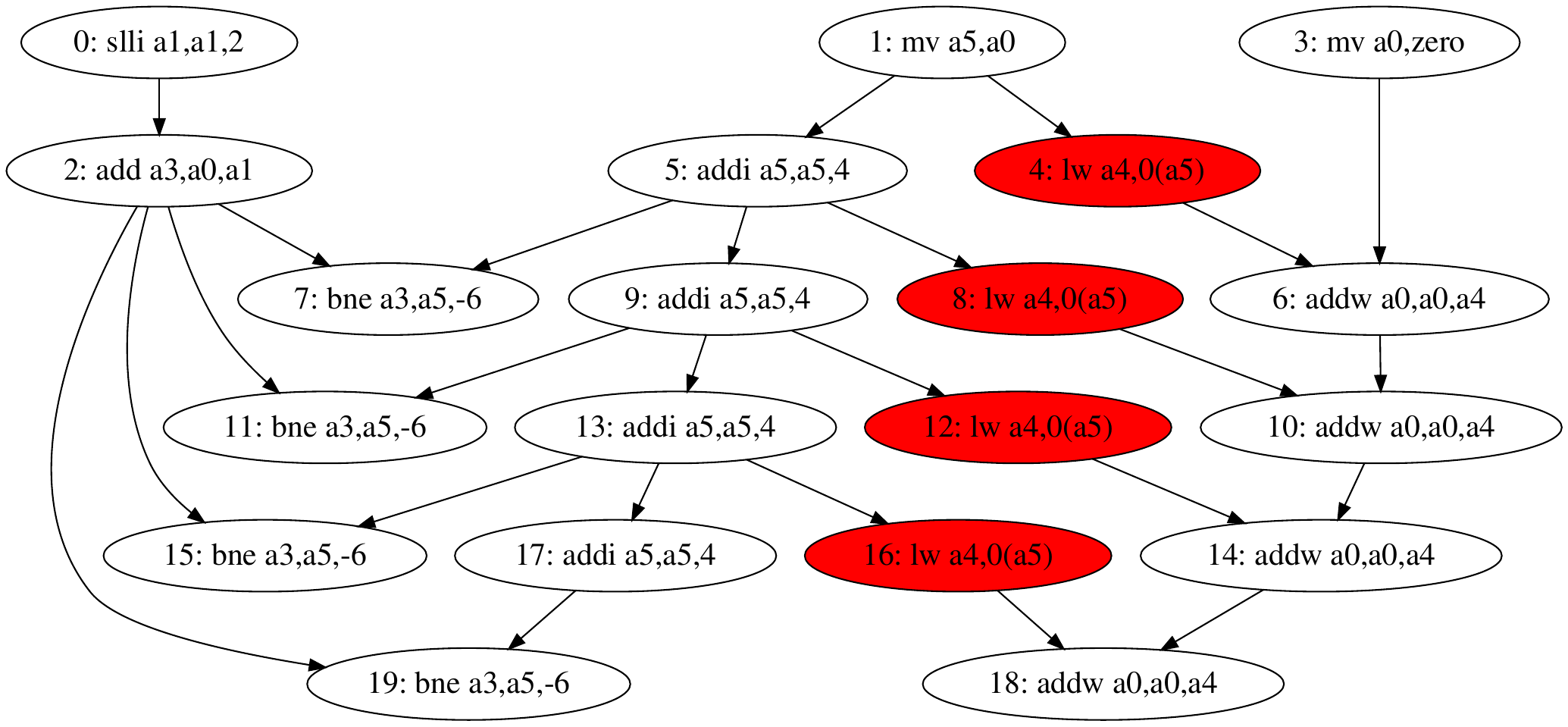}
    \caption{eDAG generated from the trace of the summation kernel for n=4. Red vertices represent memory accesses, white vertices denote other instructions. Edges represent true dependencies.}
    \label{fig:example-edag-sum-4}
\end{figure}

Fig \ref{fig:example-edag-sum-4} demonstrates the eDAG generated from the summation kernel trace with a 4-item input array. The eDAG shows that register \texttt{a0} stores \texttt{sum}, and vertices on the right add each array item to the sum. On the left, register \texttt{a5} increments as the index into the array, serving as the address for the next item to load. Branch instruction vertices (e.g., 7, 11, 15) do not overwrite register or memory address values. As eDAG only considers data dependencies and ignores control flow dependencies, no other vertices depend on them.

\subsection{eDAG Analysis}
\label{sec:edag-analysis}

To achieve the primary objective of this work, which is to obtain performance metrics including the theoretical memory latency sensitivity of a program, the generated eDAGs are passed to the eDAG analyzer. Before discussing the details of the metrics, we first define an appropriate cost model.

\subsubsection{Memory Cost Model}

To mitigate the effects of the high latency and low bandwidth of memory accesses, CPUs employ two orthogonal techniques. First, multiple memory access instructions can be pipelined and executed in parallel \cite{hardware_support_golden}. This can increase the throughput of the system as long as the \emph{memory-level parallelism} is high enough. Second, caches can reduce the number of memory accesses that need to access RAM, which improves the average latency as long as the program has good \emph{locality} \cite{improved_multithreading_boothe}. We extract a metric that takes into account memory-level parallelism and locality from our eDAGs.
We assign each memory access that goes to RAM a constant access latency of $\alpha$ and we assume that $m$ memory accesses can be issued in parallel. This means that issuing $s$ memory accesses in parallel costs $\lceil \frac{s}{m} \rceil  \alpha$. In contrast, for a chain of $s$ dependent accesses to RAM, the cost is $s \alpha$. In addition to the memory cost, we consider the total computational cost of non-memory-access operations (e.g., arithmetic, and cache access) in an eDAG to be a constant $C$ that is independent of $\alpha$ and proportional to the work. A vertex that accesses RAM (and is hence a cache miss) is a \emph{memory access vertex}.

In general, we subdivide the eDAG into layers, which we define recursively: The first layer consists of memory access vertices that are not reachable from any other memory access vertices. The $i+1$-th layer consists of memory access vertices that are reachable from vertices in the $i$-th layer without going through another memory access vertex. The number of layers is the \emph{memory depth} $\mathcal{D}$ and the total number of memory access vertices is the memory work $\mathcal{W}$. Let $\mathcal{W}_i$ be the number of memory access vertices in level $i$.
Based on these variables, the memory cost $M_{m, \alpha}$ is bounded by
\begin{align}\label{eq:memory-cost}
\max\left(\mathcal{D}, \frac{\mathcal{W}}{m} \right) \alpha \leq  M_{m, \alpha} \leq \left (\frac{\mathcal{W} - \mathcal{D} }{m} + \mathcal{D} \right ) \alpha \enspace .
\end{align}
which can be obtained in similar reasoning as the work/span laws and Brent's lemma: For the lower bounds, notice that memory accesses that depend on each other must execute one after the other. Consider a path that contains the largest number of memory access vertices in the eDAG. Its length is $\mathcal{D}$, which yields the lower bound of $\mathcal{D}\alpha$. For the second lower bound, notice that at most $m$ memory accesses can occur in parallel. As there are $\mathcal{W}$ memory access vertices, the bound $\frac{\mathcal{W}}{m}$ follows. For the upper bound, observe that by definition of a layer, 
every memory access in a given layer can be issued in parallel. Hence, the memory cost to execute layer $i$ is $\left \lceil \frac{\mathcal{W}_i}{m} \right \rceil \alpha$ and the total memory cost is bounded by $\sum_{i=1}^{\mathcal{D}} \left \lceil \frac{\mathcal{W}_i}{m} \right \rceil \alpha$ over the layers. Then, the inequality follows by using the rule $\lceil \frac{n}{m} \rceil= \lfloor \frac{n-1}{m}\rfloor +1$, which holds for positive $n$ and $m$. To be more precise, Equation \ref{eq:memory-cost} describes the theoretical upper and lower bounds of the execution time of a program if its eDAG only contains memory access vertices while all other operations are ignored. Note that variables $\mathcal{D}$, $\mathcal{W}$, $C$, $M_{m, \alpha}$ and $T_{m, \alpha}$ are all functions of a given eDAG $G$. If $G$ is clear from the context, it is omitted from the expressions. Now, if we take into account the constant computation cost of non-memory-access vertices we can bound the total theoretical cost of the eDAG, $T_{m, \alpha}$, for a given $m$ and $\alpha$ as:
\begin{align}
\max\left(\mathcal{D}, \frac{\mathcal{W}}{m} \right ) \alpha + C \leq  T_{m, \alpha} \leq \left (\frac{\mathcal{W} - \mathcal{D} }{m} + \mathcal{D} \right ) \alpha + C\enspace .
\end{align}
For simplicity, this cost model ignores the interactions between memory access vertices and other instructions. Nevertheless, it still provides us with an effective estimation of the impact of both the memory access latency and the number of available memory issue slots without having to develop a complex model that considers all the intricacies of the underlying architecture.

\subsubsection{Memory Latency Sensitivity}

In mathematics, sensitivity analysis (SA) investigates how a set of $N$ input variables $x = \{ x_1, \ldots x_N \}$ influences the output $y = \{ y_1, \ldots, y_D \}$ of a function $y = g(x)$ where $g: \mathbb{R}^N \rightarrow \mathbb{R}^D$ \cite{sensitivity_analysis_borgonovo, sensitivity_analysis_saltelli}. Two approaches are commonly used: local and global SA. Global SA provides a comprehensive view of the effects of parameters in $x$, while local SA is easier to implement and less computationally demanding. However, local SA has limitations, especially for nonlinear models, leading to biased results \cite{sensitivity_analysis_could_better_saltelli, why_so_many_saltelli}. When $g$ is differentiable, derivative-based local SA can be performed by computing the partial derivative of $y$ with respect to the $i$-th input $x_i$, denoted as $S_i = \frac{\partial y}{\partial x_i}\Bigr|_{x_0}$, where $S_i$ is the sensitivity measure of $x_i$, and $x_0 \in \mathbb{R}^n$ is the fixed point for evaluating the derivative \cite{sensitivity_analysis_of_model_output_borgonovo, sensitivity_analysis_of_environmental_models_pianosi}.

To derive memory latency sensitivity based on eDAGs, we refer to the theory regarding derivative-based local SA. In essence, we can take one of the bounds of $T_{m, \alpha}$, and compute its partial derivative with respect to $\alpha$. It is evident that the derivative expresses the quantity of how much $T_{m, \alpha}$ changes as $\alpha$ varies, and can be utilized to directly gauge the memory latency sensitivity of an application. Moreover, considering that the model is linear, we also minimize the influence of bias from local SA~\cite{addressing_uncertainty_reed, why_so_many_saltelli}. Since we are interested in the worst-case performance, we opted for the upper bound of $T_{m,\alpha}$, and define the \emph{absolute memory latency sensitivity} $\lambda$ as
\begin{align}\label{eq:absolute-mem-lat-sens}
\lambda = \frac{\partial \left( \left (\frac{\mathcal{W} - \mathcal{D} }{m} + \mathcal{D} \right ) \alpha + C \right)}{\partial \alpha} = \frac{\mathcal{W} - \mathcal{D} }{m} + \mathcal{D}
\end{align}

\begin{figure}[!t]
\begin{subfigure}{.22\textwidth}
  \centering
  \includegraphics[scale=0.38]{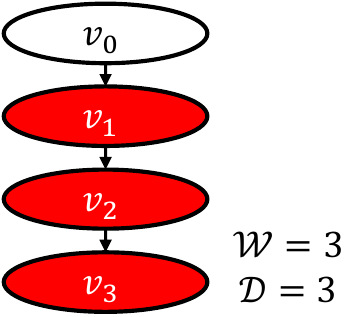}
  \caption{Example eDAG of a latency-sensitive application.}
  \label{fig:example-latency-sensitive-edag}
\end{subfigure}
\hspace{1mm}%
\begin{subfigure}{.22\textwidth}
  \centering
  \includegraphics[scale=0.38]{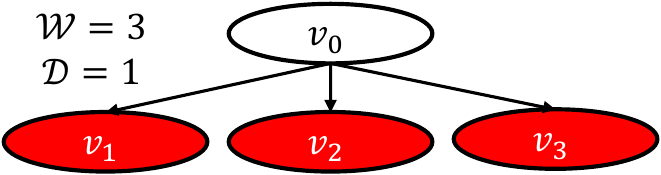}
  \caption{Example eDAG of a latency-insensitive application.}
  \label{fig:example-latency-insensitive-edag}
\end{subfigure}%
\caption{eDAGs generated from a latency-sensitive vs. latency-insensitive application, red vertices denote memory accesses.}
\label{fig:edag-comparison}
\end{figure}

Fig \ref{fig:edag-comparison} demonstrates eDAG features that distinguish a latency-sensitive application from a latency-insensitive one. eDAG $G_1$ (Fig \ref{fig:example-latency-sensitive-edag}) is more sensitive to memory latency, having memory access vertices clustered along the critical path, while $G_2$ (Fig \ref{fig:example-latency-insensitive-edag}) is more tolerant with a memory depth of 1. The effect of depth on memory latency sensitivity is constrained by the available memory issue slots, as captured by Equation \ref{eq:absolute-mem-lat-sens}. When $\alpha$ increases, $T_{m, \alpha}(G_1)$ increases by 3, while $T_{m, \alpha}(G_2)$ increases by 1. Limiting $m$ to 1 results in a cost increase of 3 for both. From this example, it can be seen that the effect of depth on the overall memory latency sensitivity is constrained by the number of available memory issue slots. This characteristic is summarized perfectly by Equation \ref{eq:absolute-mem-lat-sens}. After re-arranging, we have $\lambda = \frac{1}{m}\mathcal{W} + (1 - \frac{1}{m})\mathcal{D}$, which signifies that given a fixed $m$ and $\mathcal{W}$, $\lambda$ grows with $\mathcal{D}$. If $\mathcal{W}$ and $\mathcal{D}$ stay constant, $m$ controls the proportions of $\mathcal{W}$ and $\mathcal{D}$ in the total computation cost. The larger $m$ becomes, the more weight is given to $\mathcal{D}$ and vice-versa.

While $\lambda$ is a useful metric, it does not fully describe how an application's performance will be affected relative to a baseline. If an application is already slow, introducing additional memory access latency could lead to a comparatively larger decrease in performance on an absolute scale. However, the slowdown may not be as significant relative to its baseline performance. Conversely, adding memory latency to a fast program may only result in a minor increase in execution time on an absolute scale, but the relative impact on performance could be substantial. To this end, we formulate the \emph{relative memory latency sensitivity} $\Lambda$ of an eDAG as
\begin{align}
    \Lambda = \frac{\lambda}{\lambda \alpha_0 + C}
\end{align}
where $\alpha_0$ is the baseline latency of memory access operations. Unlike $\lambda$, $\Lambda$ is a normalized metric between 0 and 1, taking into account the percentage of a program's total computation cost attributed to memory accesses. Intuitively, it represents the relative performance change of an application with respect to a specific baseline.

\subsubsection{Bandwidth Utilization}

In addition to memory latency sensitivity, one can also approximate a program's average bandwidth utilization and visualize its data movement with the help of eDAGs. To do so, we first formulate the critical path length $T_\infty$ exactly as described in Section \ref{sec:dag-analysis}, and $w(v)$ as the amount of data moved between the CPU and the main memory in bytes when $v$ is processed. Then, under the assumption of a greedy scheduler and that an infinite number of instructions can be performed in parallel, the \emph{average bandwidth utilization} $B$ can be expressed as
\begin{align}\label{eq:relative-mem-lat-sens}
    B = \frac{\sum_{v \in V} w(v)}{T_\infty}
\end{align}
Note that $B$ should be regarded as a reference to the theoretical maximum average bandwidth that can be achieved rather than an estimate of the actual bandwidth usage.

\begin{figure}[!t]
    \centering
    \includegraphics[width=.85\linewidth]{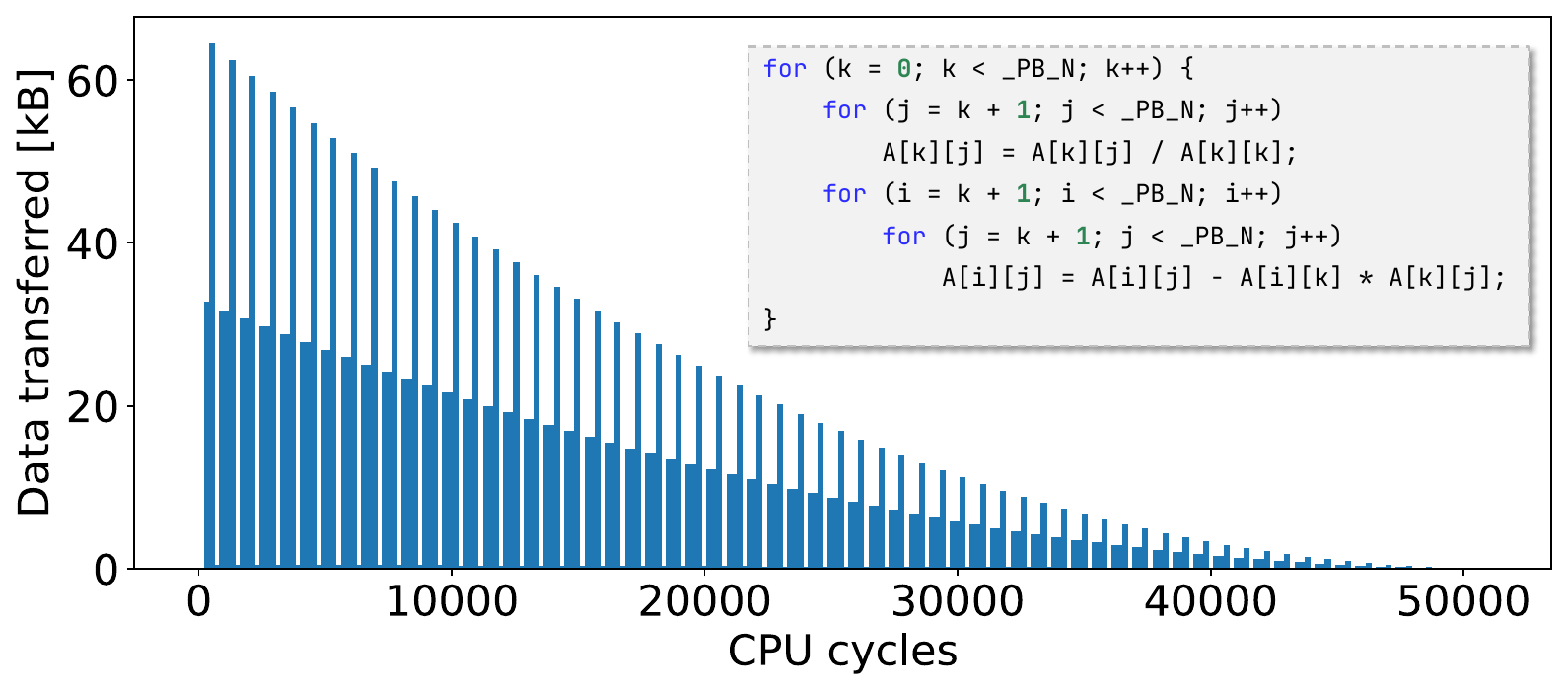}
    \caption{Data movement over time of the \emph{lu} kernel. The dataset size is 64. No cache model is used. Memory access instructions take 200 cycles, other instructions have unit costs. $\tau$ is set to 1 cycle.}
    \label{fig:data-movement-example-lu}
\end{figure}

We then define $S(v)$, and $F(v)$ as the start time and finish time of vertex $v$ respectively. Given an eDAG $G= (V, E)$, $S(v)$ and $F(v)$ can be calculated as follows
\begin{align}
    S(v) &= \begin{cases}
        0 &, \text{if }v \in I\\
        \max \left\{ F(u) | (u, v) \in E\right\} &, \text{otherwise}
    \end{cases}\\
    F(v) &= S(v) + t(v)
\end{align}
where $I$ is the set of input vertices of $G$ (i.e. vertices whose in-degree is 0), and $t$ is a predefined function that outputs the execution time of $v$. Now, we can stratify the eDAG into $\lceil \frac{T_\infty}{\tau} \rceil$ phases given a specified time interval $\tau$, and the total data movement $U_i$ within phase $i$ can be expressed as $U_i = \sum_{v \in K} w(v)$ where $K = \{ v \:|\: S(v) \leq \tau \cdot i \leq F(v) \}$ is a set containing all vertices that are being run in phase $i$. By assigning reasonable execution times to different types of instructions and adjusting the value of $\tau$, we can obtain sensible estimations of the data movement pattern of an application at various time resolutions.

Fig \ref{fig:data-movement-example-lu} shows the data movement plot generated from the trace of LU decomposition. Peaks in the diagram correspond to each iteration, aligning with the intuition behind LU decomposition. The data movement decreases as the algorithm updates the upper triangular matrix from top to bottom. This example demonstrates how eDAGs can identify hidden data bursts in a program. Moreover, it illustrates that eDAGs are not only suitable for theoretical analysis but also capable of producing practical performance metrics.

\section{Validation of EDAN}
\label{sec:edan-validation}

Despite the incorporation of an instruction cost model, EDAN by no means provides a direct estimation of programs' actual runtime. Thus, to validate the memory latency sensitivity metrics and to assess EDAN's efficacy, another approach has to be taken. The technique we opted for involves measuring the performance degradation of various applications, ranking them based on the impact of memory latency, and then comparing the applications' ranks to those acquired by analyzing their eDAGs. To gather data for the first step, we had to resort to gem5 as we lacked hardware that would easily allow artificial memory access delays to be injected.

\begin{figure}[!t]
\includegraphics[width=\linewidth]{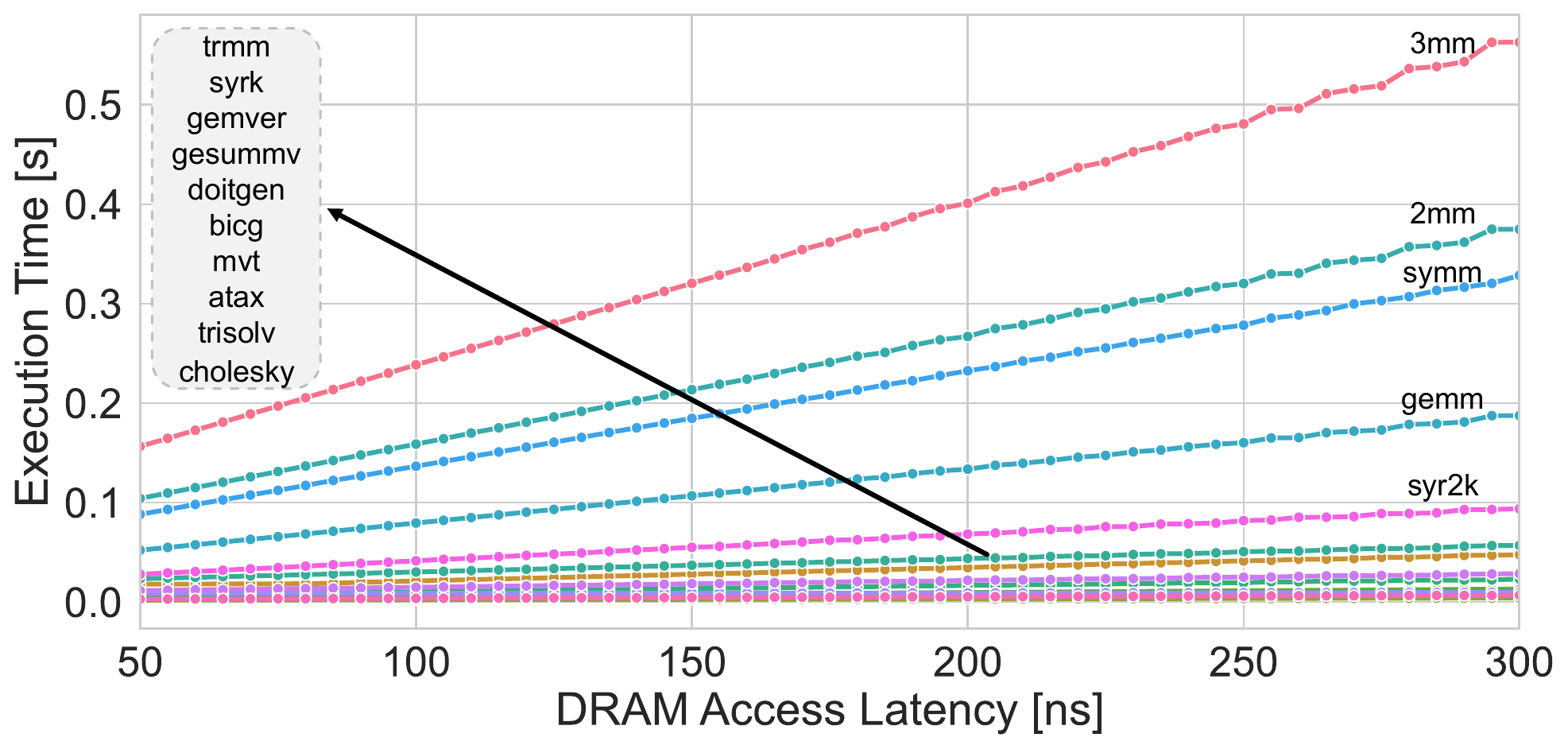}
\caption{Impact of increased memory access latency on the runtime of 15 PolyBench linear algebra benchmarks.}
\label{fig:gem5-kernels-absolute-plot}
\end{figure}

\begin{figure}[!t]
\includegraphics[width=\linewidth]{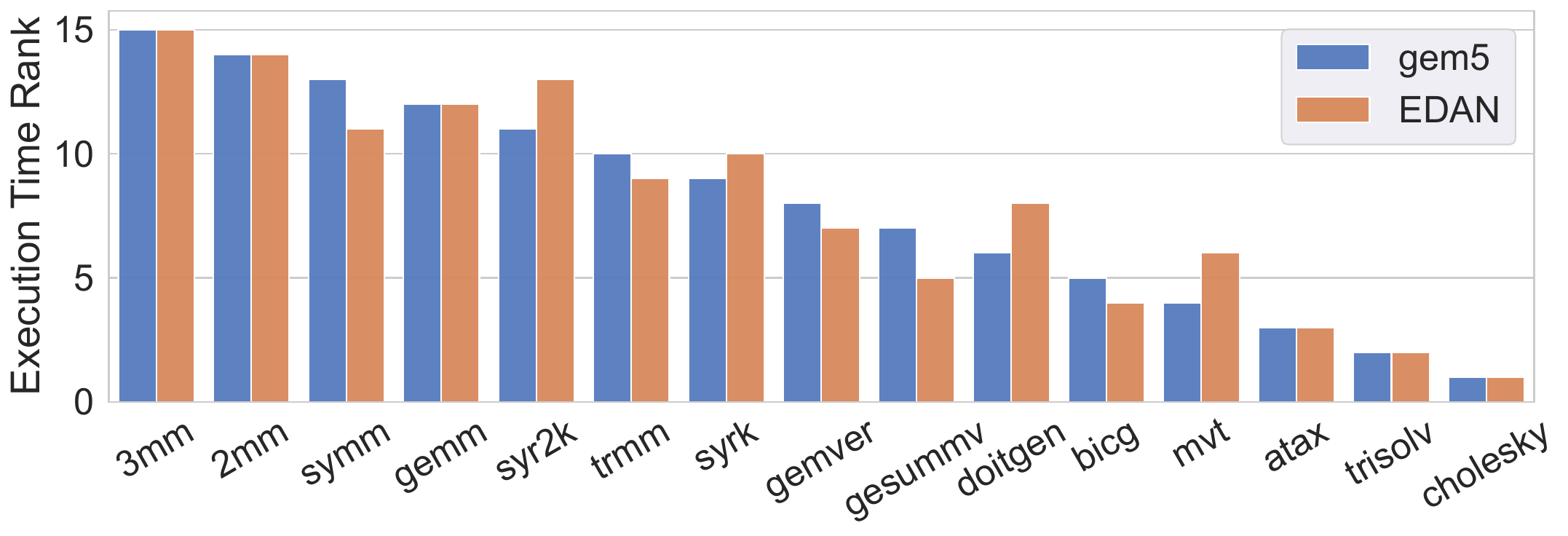}
\caption{Comparison of memory latency sensitivity rankings of benchmarks based on gem5 data and $\lambda$.}
\label{fig:absolute-mls-rank}
\end{figure}

Considering the significant latency overhead of gem5, we chose a set of linear algebra kernels with a small dataset size from PolyBench to be evaluated. PolyBench‑C bundles 30 compact kernels that capture the ``dense’’ side of scientific computing~\cite{polybench_pouchet}.  These kernels exhibit diverse memory access patterns and algorithmic motifs found within HPC applications, machine learning training, and graph processing engines. Moreover, their small size allows for complete execution within gem5 across various latency configurations.

The configuration of gem5 is as follows: SE mode, 1 GHz RiscvO3CPU with 16GM DRAM with 50ns latency, 16kB L1i and 64kB Lid caches. For simplicity, we did not attempt to emulate a multi-node system with remote memory, as performing parameter sweeps in gem5 for numerous parameters and applications will be excessively time-consuming. Instead, we varied the DRAM latency for all memory access instructions from the baseline to $300ns$ at $5ns$ increment. We used PolyBench's internal time reporting functionality to measure only the time of the computation kernel~\cite{polybench_pouchet}.

\subsection{Validation of $\lambda$}

In Fig \ref{fig:gem5-kernels-absolute-plot}, we display the runtime of 15 linear algebra benchmarks in gem5 plotted against increasing DRAM latencies. To generate the ranking, we calculated the average execution time for each across all tested latencies. We sorted them from highest to lowest, with the first kernel being the most latency sensitive. To produce the ranking with EDAN, we began by recording traces for the main computational kernels in the benchmarks, while disregarding extraneous functions like array initialization. This ensured that the code section we traced would correspond accurately with the timing data provided by gem5. We generated the eDAG for each benchmark using the same parameters for the cache model as those used in gem5. We then calculated their respective $\lambda$ value with a value of 4 for $m$ and sorted them in descending order.

Fig \ref{fig:absolute-mls-rank} compares rankings from gem5 and the $\lambda$ metric. 6 out of 15 benchmarks' rankings match perfectly with gem5's ground truth, including the two most and three least latency-sensitive kernels. For the misaligned benchmarks, the ranks differ by a maximum of 2, with an average difference of only 0.93. This demonstrates $\lambda$ as a reliable metric to compare the potential increase in execution time of multiple applications when additional memory latency is introduced. EDAN significantly enhances productivity, reducing data collection time from 24 hours in gem5 to less than an hour. It is important to note, however, that the value of $\lambda$ does not directly correspond to the magnitude of the execution time increase.

\begin{figure}[!htp]
\includegraphics[width=\linewidth]{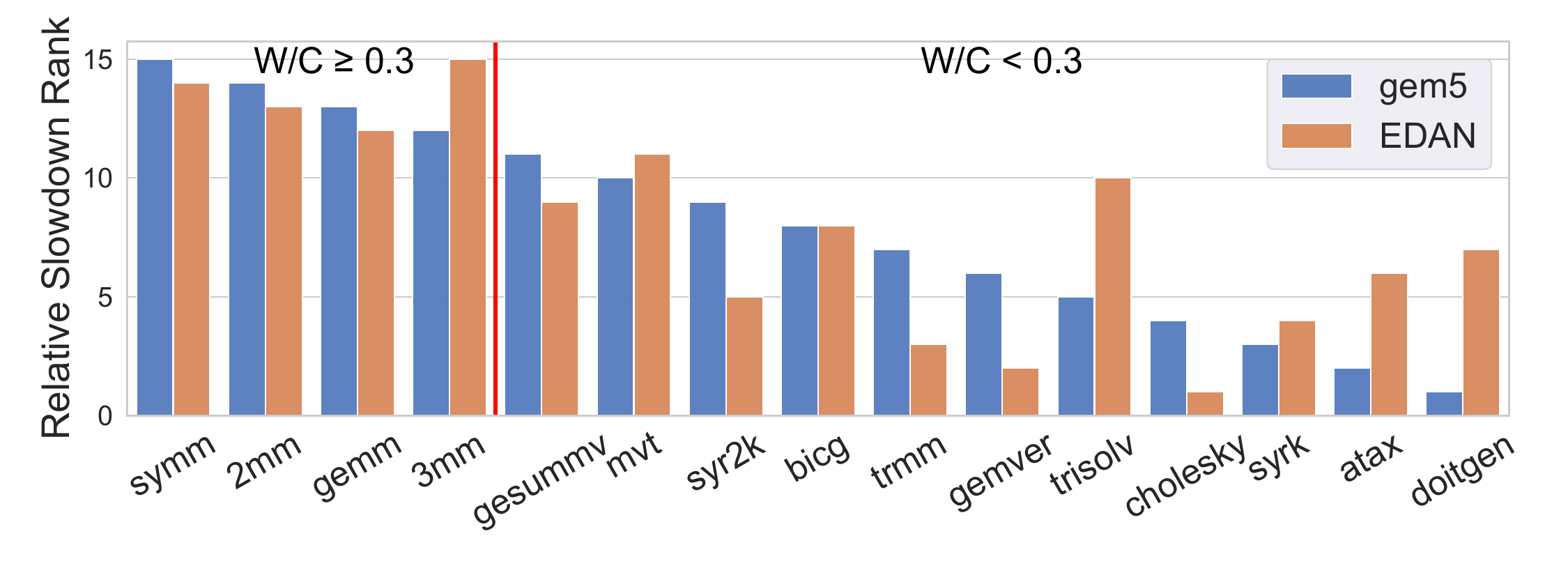}
\caption{Comparison of memory latency sensitivity rankings of benchmarks based on gem5 data and $\Lambda$.}
\label{fig:relative-mls-rank}
\end{figure}

\subsection{Validation of $\Lambda$}
\label{sec:validation-of-relative-lambda}
To test the validity of $\Lambda$, we use the same data collected with gem5. Following the same methodology outlined in the previous section, we determined the relative slowdown of each benchmark's runtime when compared to its baseline (i.e. $50ns$ DRAM latency) across all DRAM latencies. We ranked the benchmarks accordingly based on the average relative slowdown.
For all benchmarks, we chose $\alpha_0$ to be 50 and the total number of non-memory-access vertices in an eDAG to be $C$. We observed that, in this case, the actual values of $\alpha_0$ and $C$ only affect the magnitude of $\Lambda$, and do not alter the rankings of the benchmarks.

Fig \ref{fig:relative-mls-rank} presents the comparison of ranks based on the two approaches. However, the results here are noticeably poorer than those in Fig \ref{fig:absolute-mls-rank}. Specifically, only one ranking based on $\Lambda$ conforms to the ground truth, and the average discrepancy is 2.67. Nevertheless, albeit not perfectly, EDAN predicted the top 4 most latency-sensitive benchmarks using $\Lambda$. Therefore, we sought to investigate the circumstances under which $\Lambda$ would give a reasonable estimate. To do so, we computed the value of $\frac{\mathcal{W}}{C}$, the ratio of memory work to the number of non-memory-access instructions. We discovered that the top 4 kernels all have a $\frac{\mathcal{W}}{C}$ ratio larger than 0.3. Based on this finding, it can be extrapolated that in order for $\Lambda$ to provide a sensible estimate, $\frac{\mathcal{W}}{C}$ needs to be above a certain threshold. This can be attributed to the fact that our metric does not accurately model the cost of non-memory access instructions. It overlooks the interactions between memory access vertices and all other instructions, making it difficult to know when computations and memory accesses overlap or depend on each other. Since the value of $C$ cannot be computed precisely, it follows that as the proportion of memory access vertices becomes smaller, the larger the deviation between the calculated $\Lambda$ and its actual value. Despite its weakness, $\Lambda$ is still a valuable metric for identifying memory-intensive benchmarks that could benefit from performance optimization strategies such as caching or prefetching.

\section{Case Studies}

Now that we have validated our model and assessed the strengths and weaknesses of EDAN, we will proceed to apply it to a set of applications and benchmarks as case studies. This will enable us to gain an in-depth understanding of their potential memory-level parallelism and latency sensitivity.

\subsection{PolyBench-C Suite}
\label{sec:case-study-polybench}

\begin{figure}[!t]
\includegraphics[width=\linewidth]{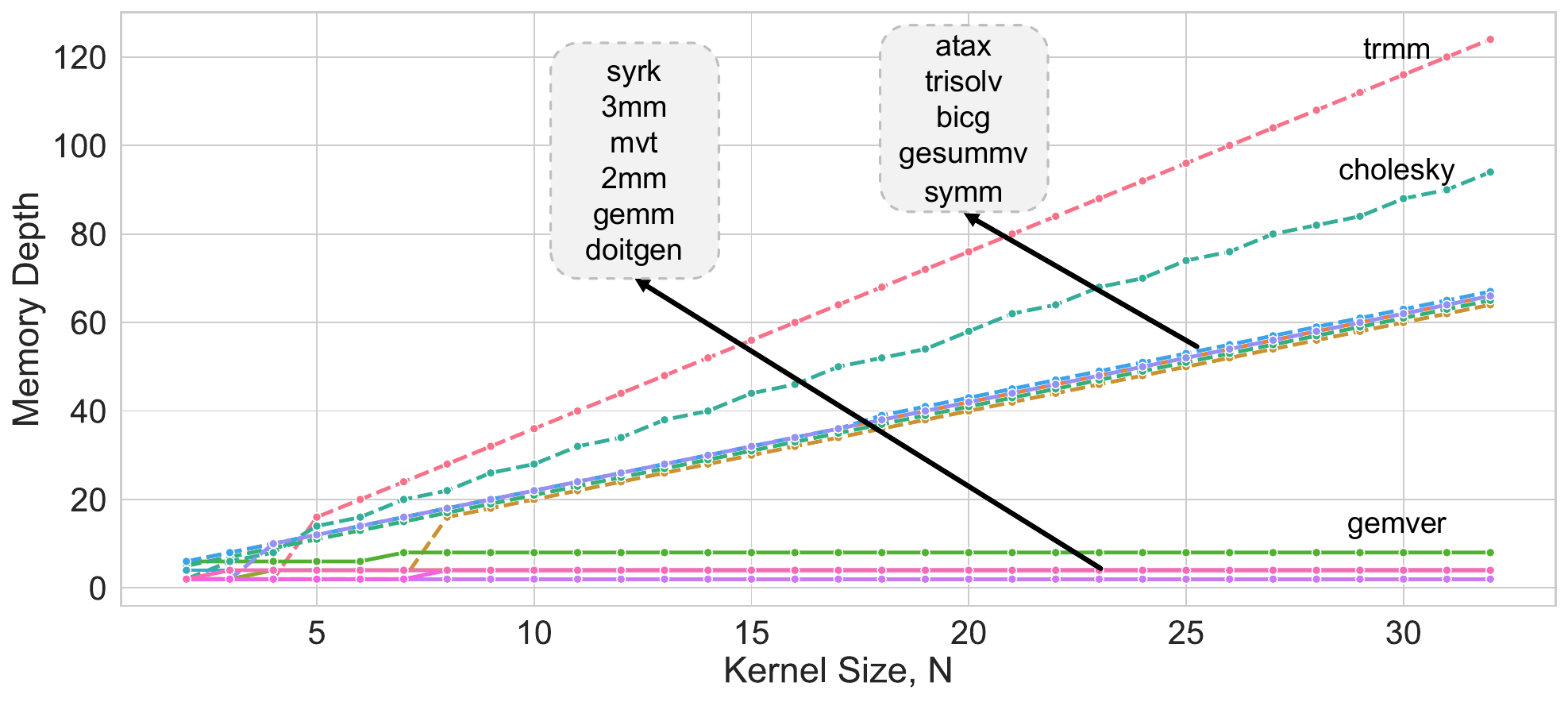}
\caption{Impact of data sizes on the memory depth $\mathcal{D}$ of PolyBench linear algebra benchmarks. Cache models were not used.
}
\label{fig:polybench-case-study}
\end{figure}

Although some experiments have already been performed on PolyBench in the previous section, analyzing $\mathcal{W}$ and $\mathcal{D}$ of individual benchmarks can still provide further insight into memory-level parallelism.
We varied the input data size $N$ for linear algebra benchmarks and investigated its impact on $\mathcal{W}$ and $\mathcal{D}$ of their eDAGs. The effect of $N$ on $\mathcal{W}$ is relatively uninformative, the relationship between them can simply be characterized by polynomial functions with different degrees according to the algorithms \cite{red_blue_pebbling_revisited_kwasniewski}. On the other hand, the connection between $N$ and $\mathcal{D}$ is more compelling. Fig \ref{fig:polybench-case-study} plots the values of $\mathcal{D}$ against $N$, and it can be seen that 8 out of 15 benchmarks have a constant memory depth despite a changing $N$. We attempted to categorize the benchmarks by the types of algorithms they perform, yet it was unsuccessful since algorithms that should belong to the same category, such as \emph{trmm} and \emph{2mm}, exhibit different behaviors for $\mathcal{D}$.

Upon closer inspection, we made the following discovery: \textbf{Data-oblivious applications exhibit constant memory depths under ideal architecture assumptions.} This is because data-oblivious applications have memory access patterns and control flows that are not dependent on the data itself \cite{data_oblivious_data_structures_mitchell}. Therefore, there are no loads that depend on each other, as in pointer chasing. If infinite registers are available, the longest chain of dependent memory accesses would involve loading a value from memory and storing it back after all dependent operations have been executed, resulting in a constant memory depth. An example would be the eDAG in Fig \ref{fig:example-edag-sum-4}, where there is only one memory access vertex along the critical path, regardless of the input size. This indicates significant memory-level parallelism. Through $\mathcal{D}$ and $\mathcal{W}$, we expose the potential memory-level parallelism in a program that is otherwise not easily detectable with a traditional work and depth model.

\begin{figure}[!htp]
\centering
\lstset{
  xleftmargin=.12\linewidth, xrightmargin=.12\linewidth
}
\begin{lstlisting}[language=C, backgroundcolor=\color{white}, basicstyle=\fontsize{6}{6}\ttfamily]
/* trmm: B := alpha*A'*B, A triangular */
for (i = 1; i < _PB_NI; i++)
  for (j = 0; j < _PB_NI; j++)
    for (k = 0; k < i; k++)
      B[i][j] += alpha * A[i][k] * B[j][k];
\end{lstlisting}
\caption{Section of source code from \emph{trmm}.}
\label{lst:trmm-3mm-source-code}
\end{figure}

Despite being data-oblivious, around half of the tested benchmarks still have a linear memory depth, which is caused by register spilling as discussed in Section \ref{sec:exposing-potential-parallelism}. To demonstrate this, we present in Fig \ref{lst:trmm-3mm-source-code} a section of the source code from \emph{trmm}. In this case, \emph{trmm} has the fastest-growing memory depth among all benchmarks. From its source code, we see that the compiler is unable to keep each \texttt{B[i][j]} in a designated register as there are too many distinct values loaded between its first and last access. For instance, when the kernel size is $4$, 15 unique values are loaded from memory between the first and last access of \texttt{B[1][0]}. Since the compiler does not keep all of them in registers, the value \texttt{B[1][0]} will be ``spilled" back to memory. This, in turn, creates extraneous dependencies between loads and stores.

\subsection{HPCG}

HPCG (High-Performance Conjugate Gradient) is a benchmark for ranking computer systems, and it centers around solving a large sparse linear system with the pre-conditioned conjugate gradient (PCG) method \cite{toward_a_new_metric_heroux, benchmarking_sterling}. Adopted by the TOP500 list in 2014 as a complement to LINPACK, it has become the reference benchmark for sparse, irregular, and communication‑heavy workloads. Its low arithmetic intensity and latency‑sensitive halos make it an ideal stress‑test for memory‑bound codes, allowing us to demonstrate EDAN's effectiveness for the sparse applications that increasingly influence HPC system design. The benchmark consists of two main phases: the setup phase and the PCG iteration phase. The setup phase constructs the sparse matrix and the multigrid hierarchy, while the PCG iteration phase performs multiple iterations of the PCG algorithm. The version of the program was 3.1. 

To analyze the program's performance, we focused on the \texttt{CG} function in the PCG iteration phase and ignored the setup phase entirely. We chose a data size of 16 and an iteration number of 50. Tracing took approximately 35 seconds and produced a file of $5.5GB$, containing over 210 million lines of instructions. The trace file was processed on a server with Intel Xeon X7550 CPUs, 1 TB of memory, and a PERC H700 hard disk. It took around 7 hours to generate and analyze the eDAG. For comparison, under identical conditions, gem5 would require approximately 4 days to produce latency sensitivity measures. We collected performance metrics from the eDAG with various cache configurations. $m$ and $\alpha_0$ were set to be 4 and 1 respectively, and $C$ is the number of non-memory access vertices. We specified the cost of memory accesses to be 200 cycles, while all other instructions had a unit cost. The cache model was a write-through 2-way associative L1 cache with 64 bytes cache line and LRU as the eviction strategy. The results are summarized in Table \ref{tab:hpcg-data}. 

\begin{table}[!htp]
\centering
\resizebox{\linewidth}{!}{%
\begin{tabular}{@{}clllll@{}}
\toprule
Cache Size & \multicolumn{1}{c}{$\mathcal{W}$} & \multicolumn{1}{c}{$\mathcal{D}$} & \multicolumn{1}{c}{$\lambda$} & \multicolumn{1}{c}{$\Lambda$} & \multicolumn{1}{c}{B {[}GB/s{]}} \\ \midrule
No Cache   & 106151255                         & 73703                             & 26593091                      & 0.1462                        & 46.5                             \\
32 kB      & 11200012 (89.4\%)                 & 45102 (38.8\%)                    & 2833830 (89.3\%)              & 0.0112 (92.3\%)               & 8.1                              \\
64 kB      & 10833505 (89.8\%)                 & 43502 (41.0\%)                    & 2741003 (89.7\%)              & 0.0108 (92.6\%)               & 8.1                              \\ \bottomrule
\end{tabular}
}
\caption{Impact of cache sizes on the performance metrics in HPCG. The numbers in parenthesis show the percentage reduction compared to the baseline.}
\label{tab:hpcg-data}
\end{table}

One can see from the data that, in this specific scenario, caching plays a significant role in mitigating the memory latency sensitivity of HPCG. Compared with the baseline in which no cache was available, we see a reduction of around 90\% for $\mathcal{W}$ when 32kB of cache is used, which results in a substantial decrease in both $\lambda$, $\Lambda$, and the average bandwidth utilization $B$. However, increasing the cache size further leads to diminishing returns as doubling the cache size does not yield a noticeable improvement in performance metrics. This can be explained by the fact that a small dataset was used, which could likely fit within the cache, at which point, only the unavoidable cold misses remain.

\begin{figure}[!htp]
\includegraphics[width=\linewidth]{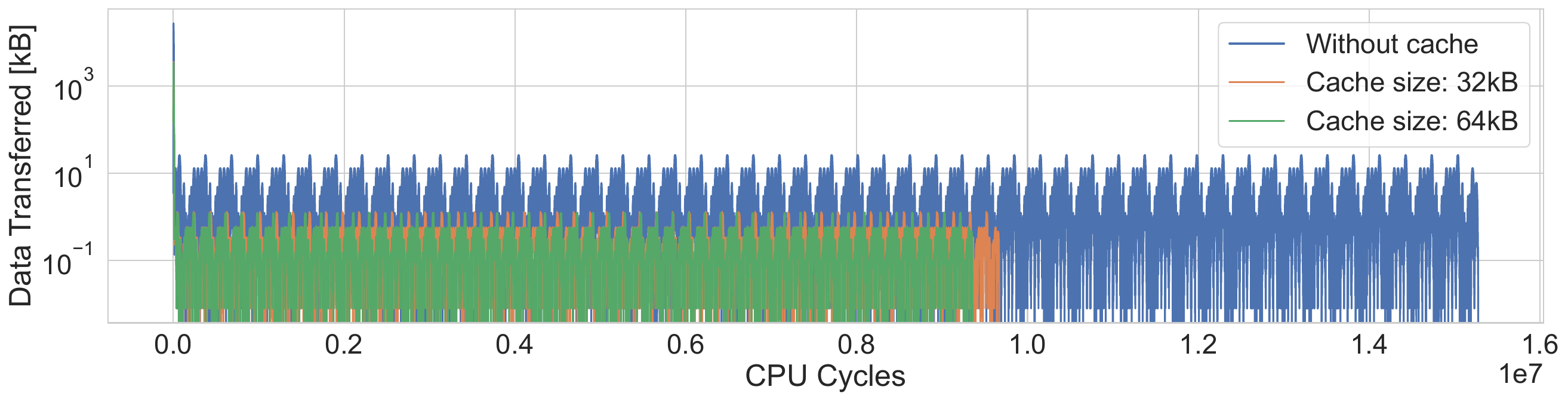}
\caption{Data movement over time of HPCG with different cache sizes. $\tau$ was set to 100 cycles.}
\label{fig:hpcg-bandwidth}
\end{figure}

We visualize the data movement over time of the three configurations in Fig \ref{fig:hpcg-bandwidth}. The pattern exhibited in the plot adheres to our intuitive understanding as a large amount of data is loaded at the start, and the repetitive small bursts of data movement coincide with each PCG iteration. There are 50 peaks in the plot, which matches perfectly the number of iterations we defined. Additionally, the impact of the cache is also visible as both the height and width of the orange and green lines are shorter compared with the baseline.

\subsection{LULESH 2.0}
To complement the synthetic kernels and sparse CG solver above, we also analyze LULESH (Livermore Unstructured Lagrangian Explicit Shock Hydrodynamics) 2.0. It is a DOE proxy application specifically developed to represent hydrodynamics codes prevalent in high-performance computing~\cite{lulesh_2_karlin, lulesh_programming_model_karlin}. Its computational kernels feature irregular mesh traversals, data-dependent memory accesses, reductions, and nearest-neighbor communication, thus testing both computational and memory-system performance under realistic conditions. Including LULESH, therefore, lets us demonstrate that EDAN can be employed for analyzing the highly irregular, latency-sensitive workloads that increasingly dominate contemporary HPC.

To better understand LULESH, we traced \texttt{LagrangeLeapFrog}, which is its kernel function, with a data size of 1000 and an iteration number of 10. The tracing process took 3.8 seconds and produced a 1.2GB file with 49 million lines of instructions. It was then processed on the same server as described in the previous section for approximately 90 minutes and computed the performance metrics with an identical set of parameters. The results are presented in Table \ref{tab:lulesh-data}.

\begin{table}[!htp]
\centering
\resizebox{\linewidth}{!}{%
\begin{tabular}{@{}clllll@{}}
\toprule
Cache Size & \multicolumn{1}{c}{$\mathcal{W}$} & \multicolumn{1}{c}{$\mathcal{D}$} & \multicolumn{1}{c}{$\lambda$} & \multicolumn{1}{c}{$\Lambda$} & \multicolumn{1}{c}{B {[}GB/s{]}} \\ \midrule
No Cache   & 18852125         & 53776          & 4753363          & 0.1370          & 13.6         \\
32 kB      & 5389537 (71.4\%) & 13083 (75.7\%) & 1357197 (71.4\%) & 0.0303 (77.9\%) & 15.8         \\
64 kB      & 5279800 (72.0\%) & 13055 (75.7\%) & 1329741 (72.0\%) & 0.0296 (78.4\%) & 15.5         \\ \bottomrule
\end{tabular}
}
\caption{Impact of cache sizes on performance metrics in LULESH.}
\label{tab:lulesh-data}
\end{table}

Compared with the data from HPCG, caching helps mitigate the memory latency sensitivity of LULESH similarly, as both $\mathcal{W}$ and $\mathcal{D}$ are decreased by more than 70\% relative to the baseline. One difference is that the majority of memory vertices are removed from the critical path, resulting in a significant reduction in $\mathcal{D}$. Hence, the critical path length $T_\infty$ is also much shorter, which leads to a slight boost in $B$.

Caching mitigates the memory latency sensitivity of LULESH, similar to HPCG, reducing both $\mathcal{W}$ and $\mathcal{D}$ by more than 70\% compared to the baseline. However, one difference is that caching removes most memory vertices from the critical path, significantly reducing $\mathcal{D}$. This results in a shorter critical path length $T_\infty$, leading to a slight boost in $B$.

\begin{figure}[!htp]

\includegraphics[width=\linewidth]{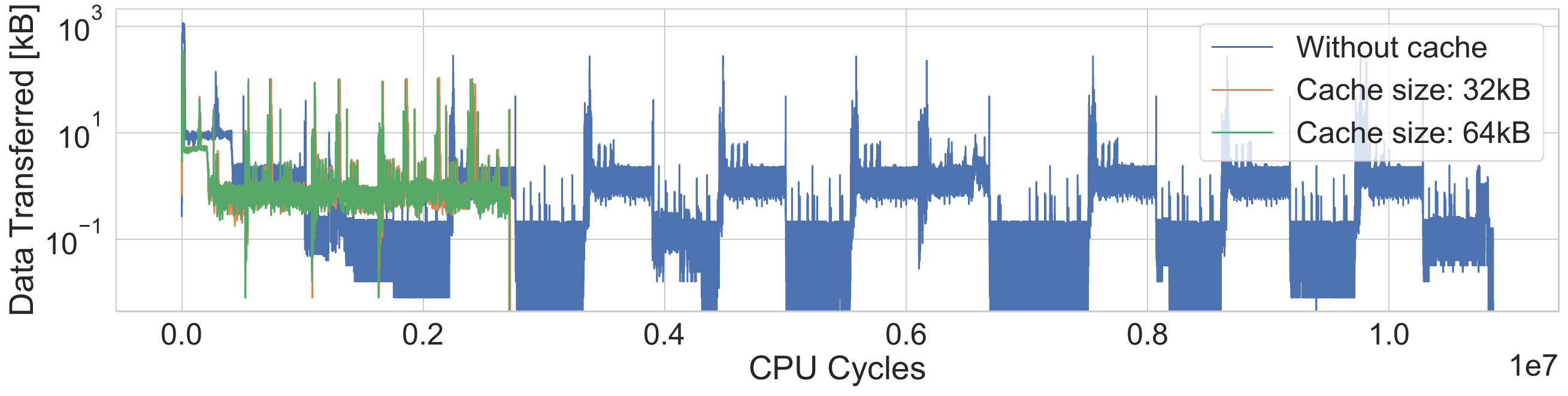}
\caption{Data movement over time of LULESH with different cache sizes. $\tau$ was set to 100 cycles.}
\label{fig:lulesh-bandwidth}
\end{figure}

Fig \ref{fig:lulesh-bandwidth} visualizes the data movement pattern of the computational kernel in LULESH, revealing its behavior during execution. The peaks in the plot indicate the start of a new time step, while the flat sections in between correspond to the calculation of nodal forces and the advancement of element quantities \cite{lulesh_programming_model_karlin}.

\section{Related Work}

\paragraph{\textbf{Performance Modeling Tools}}

Computer architects often use simulators to evaluate architectural changes, and a variety of simulators exist~\cite{gem5_binkert,sst_rodrigues2011structural,zsim_sanchez13} at different levels of detail. These tools either simulate architecture components or employ binary instrumentation~\cite{pin_luk2005} to trace events during execution and estimate latency. However, simulation tools typically require parameter sweeps to evaluate the impact of architectural changes on specific applications, leading to increased computational costs. In our work, we use the gem5 simulator primarily to validate the accuracy of our proposed model's predictions. Graph-based representations of programs are commonly employed to minimize tracing and instrumentation overhead~\cite{cabezas2014extending, railing2015collecting}.
To address these challenges, models that abstract computer details using key performance metrics (e.g., computational and memory bandwidth) have been proposed, like the original roofline model~\cite{roofline_orig}. However, these models often predict performance for simple kernels, not complex applications with diverse behaviors. Finding suitable parameters to instantiate the model can also be challenging. Other model families, such as the external memory model~\cite{balance_czechowski2011} and variations of the red-blue pebble game~\cite{redblue-pebble}, focus on differences in memory access latencies.

Another solution is to combine modeling with tracing. Tracing an application's execution allows the instantiation of the model, as shown by Cabezas and P\"{u}schel~\cite{cabezas2014extending}. They translate traced kernel execution into an execution DAG, scheduling it based on micro-architectural constraints for placement in a roofline plot. While our approach shares similarities, we directly use RISC-V binaries and combine the trace-based approach with our own model inspired by the work-span model~\cite{work_span_blelloch}. This enables us to draw conclusions about memory access parallelism effectively.

Aladdin~\cite{aladdin_shao} is a pre-RTL accelerator simulator using dynamic data dependence graphs, focusing on modeling accelerator designs and explicitly optimizing loops. In contrast, EDAN generates execution DAGs from instruction traces to capture fine-grained dependencies and inherently reveals loop concurrency by processing sequential iterations and filtering false dependencies. While both use graph representations, EDAN specifically targets quantifying memory latency sensitivity based on execution traces, complementing accelerator-focused tools like Aladdin.

The work of Alves et al.~\cite{concurrency_analysis_alves} analyzes concurrency in dynamic dataflow graphs (DDGs) with cycles to derive theoretical speed-up bounds for parallel dataflow execution models. EDAN, conversely, analyzes execution DAGs (eDAGs) from sequential instruction traces to model memory parallelism and latency sensitivity for architectural exploration, not inherent concurrency in parallel dataflow paradigms.

Graph-based Dynamic Performance (GDP)~\cite{gdp_jahre} builds a load/commit graph on chip to estimate interference‑free stall cycles at run time and drive cache‑partitioning policies.
On the other hand, EDAN is an \emph{offline} tool using complete instruction traces to generate an eDAG, analyzing intrinsic memory characteristics (parallelism, latency sensitivity) of sequential programs for architectural comparison, not runtime interference.

The work of Rakvic et al.~\cite{non_vital_loads_rakvic} classifies loads as "vital" or "non-vital" based on performance impact if delayed, finding many loads are non-vital. It proposes a "Vital Cache" storing only vital load data to improve L0 efficiency. EDAN analyzes the entire eDAG for memory parallelism and latency sensitivity metrics, rather than classifying individual loads for cache optimization.

Tune et al.~\cite{quantifying_instructions_tune} introduces a framework using graph rescheduling to precisely measure instruction criticality via slack (delay tolerance) and tautness (optimization benefit). EDAN uses eDAGs to derive aggregate metrics about memory parallelism and latency sensitivity for architectural analysis, not to quantify the criticality of every individual instruction for microarchitectural tuning.

Fields et al.~\cite{slack_fields} develop the concept of slack (instruction delay tolerance)  to guide control policies in non-uniform microarchitectures (e.g., multi-speed units). It defines slack variants and proposes hardware predictors. EDAN uses eDAG analysis for characterizing application memory behavior, not for predicting instruction slack to manage runtime resource allocation on non-uniform hardware.

The work of Srinivasan et al.~\cite{locality_srinivasan} compares memory hierarchy management based on load criticality versus traditional locality, proposing hardware to classify critical loads for optimizing caches/prefetching. It finds locality generally superior for caching. EDAN doesn't classify individual loads but analyzes the overall eDAG structure to quantify aggregate memory parallelism and latency sensitivity, focusing on application characterization rather than specific cache management policies.

\paragraph{\textbf{Memory Latency Sensitivity Analysis}}

Memory latency sensitivity analysis research can be divided into two categories: offline and online analysis. Offline analysis~\cite{locus_of_performance_domke,murphy2007effects,cabezas2014extending,lenke2021peperoni,clapp2015quantifying} focuses on understanding how workloads react to changes in the underlying machines. Some studies use architectural simulation~\cite{locus_of_performance_domke, murphy2007effects}, while others employ trace-based approaches similar to ours, but with limitations such as specific compiler toolchain requirements~\cite{cabezas2014extending} or lack of a global view of the critical path~\cite{clapp2015quantifying}. On the other hand, online analysis aims to improve system performance by dynamically changing parameters~\cite{bruno2019runtime, kim2010thread}. However, none of these studies offer a mathematical formulation of latency sensitivity, which is a unique contribution of our work.

\section{Limitations and Future Work}
EDAN is capable of efficiently producing performance metrics for a wide range of programs. Nonetheless, as a \emph{purely experimental tool}, it has a few notable limitations.

The drawbacks of the current memory cost model have already been uncovered in Section \ref{sec:edan-validation}. The lack of a more accurate CPU and scheduler model causes EDAN to mispredict the relative computation cost of non-memory access instructions, which in turn reduces the accuracy of $\Lambda$. Developing a more comprehensive model will certainly help ameliorate this discrepancy. However, this would likely introduce more computation overhead in the toolchain, and undermine the simplicity and efficiency of the current model.

As discussed extensively in Sections \ref{sec:exposing-potential-parallelism} and \ref{sec:case-study-polybench}, EDAN is constrained both by the compiler to fully expose the memory-level parallelism of a program. Techniques such as register spilling introduce extraneous dependencies and greatly hinder the discovery of memory-level parallelism. Consequently, it would be beneficial to explore the possibility of extending compilers and emulators to enable the generation and execution of code with an unlimited number of registers. 

Parallel programs are not yet supported by EDAN due to the sheer complexity of determining data dependencies in the presence of atomic operations, synchronization primitives, cache coherence protocols, and message passing. These paradigms involve intricate interactions between multiple threads and processes that create convoluted data dependencies that cannot be easily inferred from execution traces.

Since EDAN relies heavily on the execution trace, it is vulnerable to input that only triggers a particular execution path, potentially generating misleading results. Moreover, the size of input data can also impact the outcome of performance analysis depending on the compiler. Therefore, to ensure accurate and generalized results, it would be beneficial to vary the input of a program.

Although EDAN is much more efficient compared to cycle-accurate simulators, its scalability can be further improved. One approach is to store and parse traces in a binary format, reducing both the storage and computation overhead for eDAG analysis. Moreover, employing multiple processes in EDAN would enhance the processing speed of large graphs.

Currently, EDAN relies on GCC with standard extensions (i.e., MAFD) to generate binaries for the riscv64 ISA. It would be valuable to explore the impact of using different compilers and riscv64 extensions on eDAGs.

\section{Conclusion}

In this work, we present EDAN, a novel experimental toolchain that exploits the execution DAG generated from the runtime trace of a sequential program to calculate theoretical performance metrics, such as memory latency sensitivity and average bandwidth utilization. To complement the toolchain, we developed a simple yet powerful memory cost model inspired by Brent's theorem and derivative-based SA. Based on this model, we derived two metrics $\lambda$ and $\Lambda$, which can be utilized to efficiently quantify and rank the memory latency sensitivity of applications. 

By comparing our theoretical metrics with the experimental data collected from gem5, we tested the effectiveness of EDAN and understood the limitations of our model. Case studies were then conducted on several HPC benchmarks and applications, which include PolyBench, HPCG, and LULESH. Through the analysis of the performance metrics, we gained a deeper insight into the memory-level parallelism in various applications, and more importantly, we demonstrate the practicality of EDAN in the field of HPC.

As latency continues to increase in modern networks, efficient identification of application latency sensitivity is becoming increasingly crucial. With the development of EDAN, we have provided a tool to aid in this area, and we hope to inspire further advancements in this field.

\begin{acks}

The authors would like to thank Andrei Ivanov for his helpful suggestions. We also gratefully acknowledge the support of the Swiss National Supercomputing Center (CSCS) for providing computational resources.
This work was supported by the 2023 Global Research Outreach Program of the Samsung Advanced Institute of Technology (SAIT) and the Samsung R\&D Center America, Silicon Valley (SRA-SV), under the supervision of the System Architecture Laboratory (SAL). Additional funding was provided by the European Research Council (ERC) under the European Union’s Horizon 2020 research and innovation program (grant agreement PSAP, No. 101002047), as well as by the European PILOT project through the European High-Performance Computing Joint Undertaking (JU) under grant agreement No. 101034126.
This project also benefited from a donation by Intel.
ChatGPT was used for light editing of the authors' original text.
\end{acks}

\bibliographystyle{ACM-Reference-Format}
\bibliography{references}

\end{document}